\documentclass[preprint]{elsarticle}

\usepackage{amsmath}
\usepackage{amssymb}

\usepackage{graphicx}
\usepackage{subfigure}
\usepackage{url}
\usepackage{todonotes}
\usepackage{hyperref}

\usepackage[english,linesnumbered,vlined,ruled,nokwfunc]{algorithm2e}
\DontPrintSemicolon
\SetKwComment{note}{$\triangleright$ }{}
\SetFuncSty{textsc}
\SetCommentSty{textit}
\SetDataSty{texttt}
\SetKwInOut{Initial}{Initial}
\SetKwInOut{Parameter}{Param.}
\SetKwInOut{Input}{Input}
\SetKwInOut{Data}{Data}
\SetKwInOut{Output}{Output}
\ResetInOut{Output} 
\let\oldnl\nl
\newcommand{\nonl}{\renewcommand{\nl}{\let\nl\oldnl}}

\SetKw{KwAnd}{and} 
\SetKw{KwOr}{or}
\SetKw{KwXor}{xor}
\SetKw{KwNot}{not}

\newtheorem{thm}{Theorem}
\newtheorem{lem}[thm]{Lemma}
\newdefinition{defn}{Definition}
\newproof{pf}{Proof}
\newproof{pot}{Proof of Theorem \ref{thm2}}

\newcommand{\BC}{\mathrm{BC} }
\newcommand{\Bl}{\mathrm{Bl} }
\newcommand{\Br}{\mathrm{Br} }
\newcommand{\C}{\mathrm{C} }
\newcommand{\childB}{\mathrm{cB} }
\newcommand{\childS}{\mathrm{cS} }

\newcommand{\parentS}{\mathrm{pS} }
\newcommand{\SP}{\mathrm{SP} }

\newcommand{\REF}{\mathrm{ref} }

\newcommand{\sbrn}{\left[n\right]}
\newcommand{\mcsb}[2]{$\mathrm{MCS}^{\leq {#1},{#2}}$}
\newcommand{\Ss}{\Sigma_{s}}
\newcommand{\Sb}{\Sigma_{\vec b}}
\newcommand{\Edges}[1]{E({#1})}
\newcommand{\Vertices}[1]{V({#1})}

\usepackage{enumitem}
\providecommand{\varitem}{}
\makeatletter
\newenvironment{axioms}[1]
 {\renewcommand\varitem[1]{\item[\textbf{#1\arabic{enumi}\rlap{$##1$}}]%
    \edef\@currentlabel{#1\arabic{enumi}{$##1$}}}%
  \enumerate[topsep=.3em,parsep=0pt,itemsep=.3em,leftmargin=2em+\widthof{#1},label=\normalfont\textbf{#1\arabic*}, ref=#1\arabic*]}
 {\endenumerate}
\makeatother

\makeatletter
\newcommand{\DESCRIPTION@original@item}{}
\let\DESCRIPTION@original@item\item
\newcommand*{\DESCRIPTION@envir}{DESCRIPTION}
\newlength{\DESCRIPTION@totalleftmargin}
\newlength{\DESCRIPTION@linewidth}
\newcommand{\DESCRIPTION@makelabel}[1]{\llap{#1}}%
\newcommand{\DESCRIPTION@item}[1][]{%
  \setlength{\@totalleftmargin}%
       {\DESCRIPTION@totalleftmargin+\widthof{\textbf{#1 }}-\leftmargin}%
  \setlength{\linewidth}
       {\DESCRIPTION@linewidth-\widthof{\textbf{#1 }}+\leftmargin}%
  \par\parshape \@ne \@totalleftmargin \linewidth
  \DESCRIPTION@original@item[\textbf{#1}]%
}
\newenvironment{DESCRIPTION}
  {\list{}{\setlength{\labelwidth}{0cm}%
           \let\makelabel\DESCRIPTION@makelabel}%
   \setlength{\DESCRIPTION@totalleftmargin}{\@totalleftmargin}%
   \setlength{\DESCRIPTION@linewidth}{\linewidth}%
   \renewcommand{\item}{\ifx\@currenvir\DESCRIPTION@envir
                           \expandafter\DESCRIPTION@item
                        \else
                           \expandafter\DESCRIPTION@original@item
                        \fi}}
  {\endlist}
\makeatother

\makeatletter
\def\ps@pprintTitle{%
 \let\@oddhead\@empty
 \let\@evenhead\@empty
 \def\@oddfoot{%
\small%
\includegraphics[height=2em]{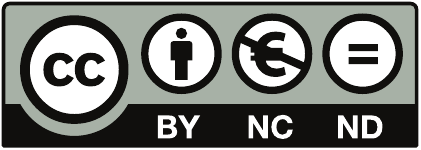}
\parbox[b][2em]{\textwidth-5.6em}{%
 \copyright~2017. This manuscript version is made available under the CC BY-NC-ND 4.0 license \url{http://creativecommons.org/licenses/by-nc-nd/4.0/}%
}
}%
 \let\@evenfoot\@oddfoot}
\makeatother

\begin{document}

\begin{frontmatter}

\title{On Maximum Common Subgraph Problems in Series-Parallel Graphs\tnoteref{iwoca,thanksDFG}}
\tnotetext[iwoca]{An extended abstract of this work appeared in \cite{Kriege2014aIWOCA}.}
\tnotetext[thanksDFG]{This work was supported by the German Research Foundation (DFG), priority programme ``Algorithms for Big Data'' (SPP 1736).}

\author{Nils~Kriege}
\ead{nils.kriege@tu-dortmund.de}

\author{Florian~Kurpicz}
\ead{florian.kurpicz@tu-dortmund.de}

\author{Petra~Mutzel}
\ead{petra.mutzel@tu-dortmund.de}

\address{Technische Universit\"at Dortmund, Department of Computer Science\\Chair of Algorithm Engineering (LS11), 44221 Dortmund, Germany}

\begin{abstract}
The complexity of the maximum common connected subgraph problem in partial $k$-trees is still not fully understood. Polynomial-time solutions are known for degree-bounded outerplanar graphs, a subclass of the partial $2$-trees. On the other hand, the problem is known to be \textbf{NP}-hard in vertex-labeled partial $11$-trees of bounded degree. We consider series-parallel graphs, i.e., partial $2$-trees. We show that the problem remains \textbf{NP}-hard in biconnected series-parallel graphs with all but one vertex of degree $3$ or less. 
A positive complexity result is presented for a related problem of high practical relevance which asks for a maximum common connected subgraph that preserves blocks and bridges of the input graphs. We present a polynomial time algorithm for this problem in series-parallel graphs, which utilizes a combination of BC- and SP-tree data structures to decompose both graphs.
\end{abstract}


\end{frontmatter}

\section{Introduction}
Finding a maximum common connected subgraph (MCS) of two input graphs is an important task in many application domains like pattern recognition and cheminformatics~\cite{Schietgat2008}. The problem is well known to be \textbf{NP}-hard. However, practically relevant graphs, e.g., derived from small molecules, often have small treewidth~\cite{Horvath2010}. Hence, it is highly relevant to develop polynomial time algorithms for tractable graph classes and to clearly identify graph classes where MCS remains \textbf{NP}-hard. For the related subgraph isomorphism problem such a clear demarcation for partial $k$-trees is known. Subgraph isomorphism is solvable in polynomial time in partial $k$-trees if the smaller graph either is $k$-connected or has bounded degree~\cite{Matousek1992,Gupta1994}. However, it is \textbf{NP}-complete when the smaller graph is not $k$-connected or has more than $k$ vertices of unbounded degree~\cite{gupta96}. MCS is at least as hard as subgraph isomorphism; two recent results show that it actually is considerably harder: Akutsu~\cite{Akutsu2013} has shown that MCS is \textbf{NP}-hard in vertex-labeled partial $11$-trees of bounded degree. Furthermore, it was believed that the problem of finding a maximum common $k$-connected subgraph of $k$-connected partial $k$-trees ($k$-MCS) can be solved with the same technique that was successfully used for subgraph isomorphism. Recently, it was shown that these techniques are insufficient even for series-parallel graphs~\cite{Kriege2014a}. However, for this class of graphs a new approach was devised, which employs SP-trees to re\-present the series parallel composition of the input graphs. Further polynomial time algorithms were proposed for MCS in almost trees and outerplanar graphs of bounded degree~\cite{Akutsu1993,Akutsu2012}.

Motivated by the fact that even subgraph isomorphism is \textbf{NP}-hard when the smaller graph is a tree and the other is outerplanar~\cite{Syslo1982}, a problem variation referred to as block-and-bridge preserving MCS (BBP-MCS) was considered~\cite{Schietgat2007,Schietgat2008,Schietgat2013}. Here, the common connected subgraph is required to inherit the structure of blocks, i.e., maximal biconnected subgraphs, and bridges of the input graphs, which renders efficient algorithms for outerplanar graphs possible~\cite{Schietgat2007}. Notably, BBP-MCS yields meaningful results for molecular graphs in practice and even compares favorably to the solutions obtained by ordinary MCS in empirical studies~\cite{Schietgat2008,Schietgat2011}.

\paragraph{Our Contribution} On the theoretical side, we prove that finding an MCS of two biconnected series-parallel graphs, i.e., partial $2$-trees~\cite{Brandstadt1999}, with all but one vertex of degree bounded by $3$ is \textbf{NP}-hard. We obtain this result by a polynomial-time reduction of the \emph{Numerical Matching with Target Sums} problem. Furthermore, we consider BBP-MCS in series-parallel graphs and propose a polynomial time solution, thus, generalizing the known result for outerplanar graphs. Employing BC- and SP-tree decompositions of the input graphs allows us to identify subproblems closely related to $k$-MCS, $k \in \{1,2\}$. We make use of a classical approach for the maximum common subtree problem~\cite{Matula1978}, i.e., $1$-MCS, and a recently proposed algorithm for $2$-MCS~\cite{Kriege2014a} to obtain our result. Our approach yields a running time of $\mathcal{O}(n^6)$ in series-parallel and $\mathcal{O}(n^5)$ in outerplanar graphs, where $n$ is the maximum number of vertices in one of the input graphs.

\section{Preliminaries} 
\label{sec:preliminaries}
	We consider \emph{simple} graphs, i.e., a graph $G$ without loops and multiple edges. We denote the finite set of \emph{vertices} by $\Vertices{G}$ and the finite set of \emph{edges} by $\Edges{G}$. A graph $G^{\prime}$ is a \emph{subgraph} of $G$, denoted by $G^{\prime}\subseteq G$, if $\Vertices{G^{\prime}}\subseteq\Vertices{G}$ and $\Edges{G^{\prime}}\subseteq\Edges{G}$. 
   A subgraph $G^{\prime} \subseteq G$ is said to be \emph{proper} if $G^{\prime} \neq G$ and we write $G' \subset G$.
   A subgraph $G \subseteq H$ is called \emph{maximal} regarding a property if $G$ itself has the property and there is no graph $G^{\prime}$ which has the property and satisfies $G \subset G^{\prime} \subseteq H$.
   For two graphs $G=(V,E)$ and $G^{\prime}=(V',E')$, we denote by $G \cup G^{\prime}$ the graph $(V \cup V^{\prime}, E \cup E^{\prime}$). For short, we write $G \cup \left\{ v\right\}$ and $G \cup \left\{ e\right\}$ to denote the union with a graph consisting of a single vertex $v$ and a single edge $e$ with its two endpoints, respectively.
   A graph is \emph{connected} if there is a path between any two vertices. Each maximal connected subgraph $G^{\prime}\subseteq G$ is called a \emph{connected component}. Let $V\subseteq\Vertices{G}$, then $G[V]$ denotes the \emph{induced subgraph} $G^{\prime}\subseteq G$ with $\Vertices{G^{\prime}}=V$ and $\Edges{G^{\prime}}=\left\{ (u,v)\in V\times V\colon (u,v)\in\Edges{G} \right\}$. A set $S \subseteq \Vertices{G}$ is called \emph{$\vert S\vert$-separator} or \emph{separator} of a connected graph $G$ if $G \setminus S := G[\Vertices{G} \setminus S]$ consists of at least two connected components. If $S=\left\{ v \right\}$ is a separator then $v$ is called \emph{cutvertex}. A separator $S$ is said to \emph{separate} two vertices $a,b \in V(G)$ if $a$ and $b$ are in different connected components of $G \setminus S$. A separator $S$ of $G$ is called \emph{minimal} if there are vertices $a,b \in V(G)$ that are separated by $S$ and there is no separator $S' \subset S$ that separates $a$ and $b$. A graph $G$ with $|V(G)| > k$ is called \emph{$k$-connected} if there is no $j$-separator of $G$ such that $j<k$ and \emph{biconnected} if $k=2$. We define $[n]:=\left\{ 1,\dots, n \right\}$ for all $n\in\mathbb{N}$. A sequence of distinct vertices $(v_0,v_1,\dots,v_n)$ such that $(v_{i-1},v_i) \in E(G)$ for all $i\in[n]$ is called \emph{path}. The vertices and the edges connecting consecutive vertices are said to be \emph{contained} in the path. If all but the first and the last vertex are distinct, i.e., $v_n=v_0$, the sequence is called \emph{cycle}. The \emph{length} of a path or cycle is the number of edges contained in it. A \emph{chord} is an edge $e$ connecting two vertices of a cycle that does not contain the edge $e$.

    A graph $G$ is \emph{bipartite} if there are two disjoint sets $U,U^{\prime}\subseteq\Vertices{G}$ such that $U\cup U^{\prime}=\Vertices{G}$ and for all $(u,v)\in\Edges{G}$ neither $u,v\in U$ nor $u,v\in U^{\prime}$. A \emph{matching} in $G$ is a set of edges $M\subseteq\Edges{G}$ such that $u=u^{\prime}\iff v=v^{\prime}$ for all $\left((u,v),(u^{\prime},v^{\prime})\right)\in M\times M$. In a \emph{weighted} graph $G$, each edge $e\in\Edges{G}$ is associated with a weight $w(e)\in\mathbb{R}\cup\left\{ -\infty \right\}$. The \emph{maximum weighted bipartite matching} problem (MwbM) asks for the maximum weight of a matching in a weighted bipartite graph and is solvable in $\mathcal{O}(n^{3})$, e.g., with the Hungarian method \cite{Kuhn55}.
	
    \begin{figure}[t!]
		\null\hfill
		\subfigure[]{	
			\includegraphics{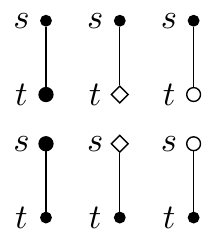}
			\label{sfig:sp_1}
		}\hfill
		\subfigure[]{	
			\includegraphics{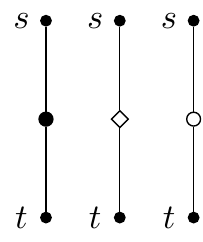}
			\label{sfig:sp_2}
		}\hfill
		\subfigure[]{	
			\includegraphics{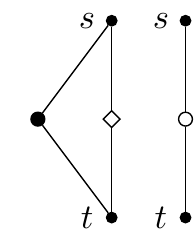}
			\label{sfig:sp_3}
		}\hfill
		\subfigure[]{	
			\includegraphics{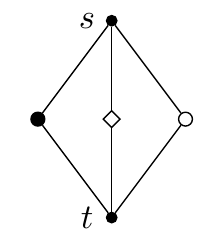}
			\label{sfig:sp_4}
		}\hfill\null
		\caption{\subref{sfig:sp_1} A set of six $K^{s,t}_2$, \subref{sfig:sp_2} the set after an $S$-operation is applied to the $s$- and $t$- nodes of the same shape, \subref{sfig:sp_3} the set after a $P$-operation is applied to the first two components and finally \subref{sfig:sp_4} the set after a $P$-operation is applied to the remaining pair of $s$- and $t$-nodes resulting in a series-parallel graph.}
		\label{fig:example_of_sp_operations}
	\end{figure}

    A graph is \emph{complete} if there is an edge between any two vertices. $K_n$ denotes the complete graph with $n$ vertices and $K^{s,t}_{2}$ denotes an instance of the $K_{2}$ where one vertex is called \emph{s}- and the other \emph{t}-vertex. A graph is \emph{series-parallel} if each maximal biconnected subgraph can be constructed starting with a finite set of $K^{s,t}_{2}$ by performing a sequence of the following two operations, see Figure~\ref{fig:example_of_sp_operations}.
    \begin{DESCRIPTION}
        \item[$S$-Operation] Merge the \emph{s}-vertex of one connected component with the \emph{t}-vertex of a different connected component. The vertex created by merging remains unnamed.
        \item[$P$-Operation] Given two different connected components of the set, merge both \emph{s}-vertices with each other and both \emph{t}-vertices with each other. The resulting vertices are called \emph{s}- and \emph{t}-vertex, respectively.
    \end{DESCRIPTION}
    By definition, series-parallel graphs are at most biconnected and equivalent to partial $2$-trees, i.e., graphs with treewidth at most $2$~\cite{Brandstadt1999}.

	Two graphs $G$ and $H$ are said to be \emph{isomorphic}, written $G \simeq H$, if there is a bijection $\phi\colon \Vertices{G}\to \Vertices{H}$, such that $(u,v)\in E(G)\iff (\phi(u),\phi(v))\in E(H)~\forall u,v\in \Vertices{G}$. A graph $H$ is \emph{subgraph isomorphic} to $G$, if $H$ is isomorphic to a subgraph of $G$. We say $u$ is \emph{mapped} to $v^{\prime}$ if $\phi(u)=v^{\prime}$. 
   A \emph{common (induced) subgraph isomorphism} $\phi$ between $G$ and $H$ is an isomorphism between induced subgraphs $G[R]$ and $H[S]$ obtained for two subsets $R\subseteq \Vertices{G}$ and $S\subseteq \Vertices{H}$; in this case the graph $G[R] \simeq H[S]$ is called \emph{common (induced) subgraph} of $G$ and $H$.
   A common subgraph isomorphism $\phi$ is called \emph{maximum} if there is no common subgraph isomorphism $\phi'$ with $|\textrm{dom}(\phi')|>|\textrm{dom}(\phi)|$, where $\mathrm{dom}(\phi)$ denotes the \emph{domain} of $\phi$. A common subgraph is called \emph{maximum} if there is no common subgraph containing more vertices. The problem we consider is defined as follows.
	\begin{defn}[Maximum Common Subgraph Problem (MCS)]
		Given\\ two input graphs $G$ and $H$, return the number of vertices in a maximum common connected induced subgraph.
	\end{defn}

	Please notice, that MCS can denote both the problem and a subgraph. In the following we assume that the input graphs are connected series-parallel graphs and common subgraphs must be induced connected subgraphs of both input graphs.


\section{MCS in Series-Parallel Graphs with Bounded Degree} 
\label{sec:series-parallel_graphs_with_bounded_degree}

	In this section, we consider MCS where both input graphs are biconnected and have degree at most 3 for all but one vertex (\mcsb{3}{1}). We prove that this problem is \textbf{NP}-hard and improve the result for subgraph isomorphism that, transferred to MCS, states that \mcsb{4}{2} is \textbf{NP}-hard~\cite{gupta96}. 

	Since the running time of an algorithm is given with respect to the size of the input, a reasonable encoding is demanded, e.g., the integer $n$ can be encoded in $\log n$ bits. An \textbf{NP}-complete problem may no longer be \textbf{NP}-complete if the instances are encoded in unary. \emph{Strongly} \textbf{NP}-complete problems are \textbf{NP}-complete even if the input is encoded in unary~\cite{garey78}. Hence, even the values of numbers can be used in a polynomial-time reduction. To prove that \mcsb{3}{1} is \textbf{NP}-hard we show that there is a polynomial-time reduction from the following problem which is \textbf{NP}-complete in the strong sense~\cite{garey79}.

	\begin{defn}[Numerical Matching with Target Sums (NMwTS)]~\\
		Given two disjoint sets $X$ and $Y$ of equal size $\left\vert X\right\vert=\left\vert Y\right\vert=n$, a value function $s\colon X\cup Y\to\mathbb{Z}^+$ associating each element in $X$ and $Y$ with positive integer value and a vector $\vec b=\langle b_1,b_2,\dots,b_n\rangle$ with $b_i\in\mathbb{Z}^+$ for all $i\in\sbrn$. NMwTS asks whether $X\cup Y$ can be partitioned into disjoint sets $A_1,A_2,\dots,A_n$ such that each set contains one element from $X$ and one element from $Y$ with $\sum_{a\in A_i}s\left(a\right)=b_{i}$ for all $i\in\sbrn$?
	\end{defn}

	\subsection{Construction of the Polynomial-Time Reduction} 
	\label{sub:the_graphs_for_a_polynomial_time_reduction}
		For an instance $(X,Y,s,\vec b)$ of NMwTS we construct two graphs, which represent the values of the elements in $X,Y$ (graph $G$) and $\vec b$ (graph $H$) in a way such that the number of vertices in an MCS of $G$ and $H$ indicates whether there is a numerical matching in the NMwTS instance. Let $\Ss:=\sum_{z\in X\cup Y}s(z)$ and $\Sb:=\sum_{i=1}^{n}b_{i}$. First we define some gadgets which are then used to construct the graphs. The gadget $B^{v}_{w}$ denotes a cycle with $2\Ss + 2$ vertices such that both paths from $v$ to $w$ have length $\Ss+1$. Next, the gadget $C^{v}_{w,w^{\prime}}$ is the graph $B^{v}_{v^{\prime}}$ with additional vertices $w, w^{\prime}$ called \emph{anchor vertex} and \emph{prime anchor vertex}, respectively, and additional edges $\left(v^{\prime},w^{\prime}\right),\left(w^{\prime},w\right)$, see Figure~\ref{sfig:Cab}. The gadget $D^{v}_{w,w^{\prime}}$ is an extension of $C^{v}_{w,w^{\prime}}$ with two chords such that it is still outerplanar and there are two edge disjoint paths of length $5$ from $v$ to $w$, see Figure~\ref{sfig:Dab}.\footnote{If an instance of NMwTS does not allow the construction of the $D^{v}_{w}$ gadgets, all values and the vector $\vec b$ are multiplied by $3$.} The gadget $K^{v}_{w}$ is a $K_{3}$, where two vertices are denoted by $v$ and $v^{\prime}$, with an additional vertex $w$ and an edge $\left(v^{\prime},w\right)$, see Figure~\ref{sfig:Kab}. Last, $P^{v}_{w}$ is a path of length $2$, where the vertices of degree $1$ are denoted by $v$ and $w$, see Figure~\ref{sfig:Pab}. We construct the graphs $G$ and $H$ as follows:
		\begin{figure}[t]
			\null\hfill
			\subfigure[$C^{v}_{w,w^{\prime}}$]{	
				\includegraphics{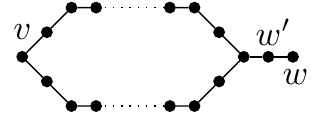}
				\label{sfig:Cab}
			}\hfill
			\subfigure[$D^{v}_{w,w^{\prime}}$]{	
				\includegraphics{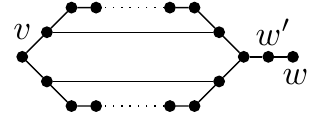}
				\label{sfig:Dab}
			}\hfill
			\subfigure[$K^{v}_{w}$]{	
				\includegraphics{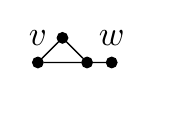}
				\label{sfig:Kab}
			}\hfill
			\subfigure[$P^{v}_{w}$]{	
				\includegraphics{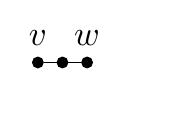}
				\label{sfig:Pab}
			}\hfill\null
			\caption{Types of gadgets used to construct the graphs $G$ and $H$.}
			\label{fig:example_of_c_d_k_and_p}
		\end{figure}

		\begin{align}
			B_{G}=	&	\bigcup_{i=1}^{n}\left(C^{\bar x}_{c_{i},c_{i}^{\prime}}\cup D^{\bar x}_{c_{i+n},c_{i+n}^{\prime}}\right)\cup\bigcup_{i=1}^{n-1}\left(P^{c_{i}^{\prime}}_{c_{i+1}^{\prime}}\right)\label{eq:base-gadget_G}\\
			B_{H}=	&	\bigcup_{i=1}^{n}\left(C^{\bar y}_{c_{i},c_{i}^{\prime}}\cup D^{\bar y}_{c_{i+n},c_{i+n}^{\prime}}\right)\cup\bigcup_{i=n+1}^{2n-1}\left(P^{c_{i}^{\prime}}_{c_{i+1}^{\prime}}\right)\label{eq:base-gadget_H}\\
			G=	&\bigcup_{i=1}^{n}\left(K^{c_{i}}_{\bar x_{1,i}}\cup K^{c_{i+n}}_{\bar y_{1,i}}\cup\bigcup_{j=2}^{s\left(x_i\right)}K^{\bar x_{j-1,i}}_{\bar x_{j,i}}\cup\bigcup_{j=2}^{s\left(y_i\right)}K^{\bar y_{j-1,i}}_{\bar y_{j,i}}\cup P^{\bar x_{s\left(x_{i}\right)}}_{\bar y_{s\left(y_{i}\right)}}\right)\cup B_{G}\label{eq:graph_g}\\
			H=	&\bigcup_{i=1}^{n}\left(K^{c_{i}}_{v_{1,i}}\cup K^{v_{b_{i},i}}_{c_{i+n}}\cup\bigcup_{j=2}^{b_i}K^{v_{j-1}}_{v_{j}}\right)\cup B_{H}\label{eq:graph_h}
		\end{align}

		\begin{figure}[ht!]
			\centering
			\includegraphics[]{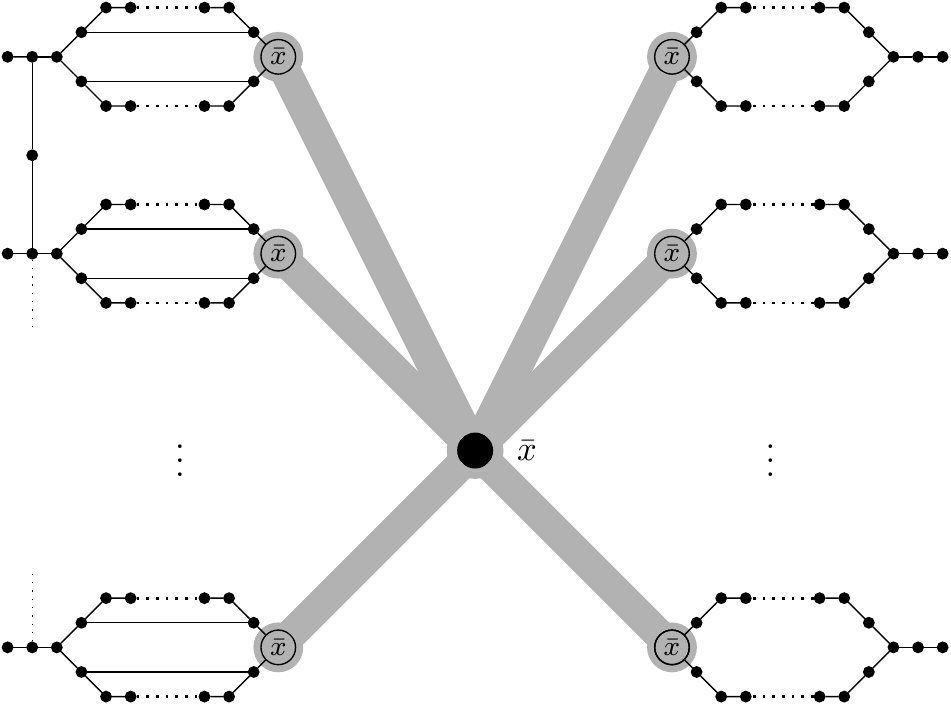}
			\caption{Base-gadget of $G$, where all vertices connected by the thick gray line represent the same vertex, i.e., vertex $\bar x$ with unbounded degree. There are $n$ gadgets $D^{u}_{v}$ on the left of $\bar x$ with the prime anchor vertices connected by $n-1$ anchor paths and $n$ gadgets $C^{u}_{v}$ on the right.}
			\label{fig:base_gadget_G}
		\end{figure}
		The graphs $B_G$ and $B_H$ (see Eq.~\ref{eq:base-gadget_G} and \ref{eq:base-gadget_H} and for the former also Figure~\ref{fig:base_gadget_G}) are called the \emph{base-gadget} of $G$ and $H$, respectively. Each of these base-gadgets consists of $n$ chordless cycles and $n$ cycles with chords, i.e., $C^{\bar x}_{c_{i}}$ and $D^{\bar x}_{c_{i+n}}$ for $i\in\left[n\right]$ in $B_G$ and $C^{\bar y}_{c_{i}}$ and $D^{\bar y}_{c_{i+n}}$ for $i\in\left[n\right]$ in $B_H$. Hence, the base-gadgets of $G$ and $H$ contain one vertex with unbounded degree named $\bar x$ and $\bar y$, respectively. The difference between the base-gadgets is that in $B_G$ there are paths of length $2$ (see $P^{v}_{w}$ in Eq.~\ref{eq:base-gadget_G} and \ref{eq:base-gadget_H}) between the prime anchor vertices of the cycles containing chords and in $B_H$ these paths are between the prime anchor vertices of the chordless cycles. Due to these paths the resulting graphs are biconnected as we will prove in Lemma~\ref{lem:biconnected}. We call the paths between the prime anchor vertices \emph{anchor paths}.

\begin{figure}[ht!]
			\centering
			\includegraphics[scale=1.1]{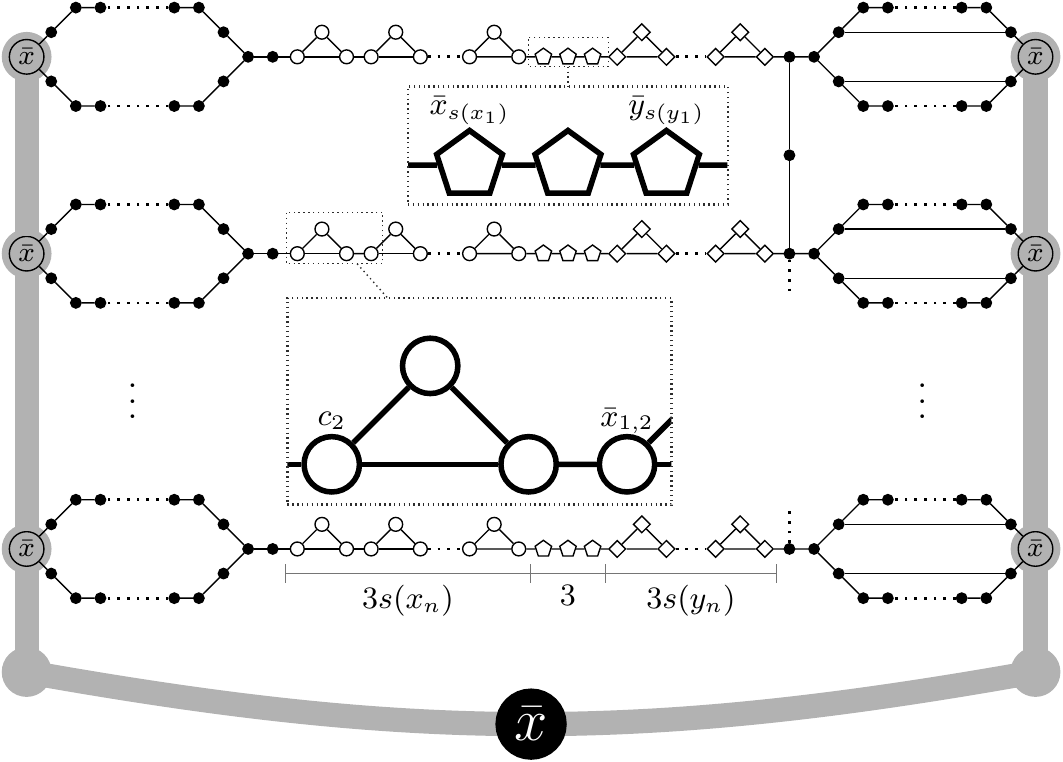}
			\caption{Graph $G$ containing the base-gadget as illustrated in Figure~\ref{fig:base_gadget_G}. All vertices connected by the thick gray line represent the same vertex $\bar x$. Circles represent $x$-paths, diamonds represent $y$-paths, where the $i$-th $x$- and $y$-path contains $3s(x_i)$ and $3s(y_i)$ vertices, respectively. Pentagons represent the separation paths containing $3$ vertices.}
			\label{fig:graph_g}
		\end{figure}

\begin{figure}[ht!]
			\centering
			\includegraphics[scale=1.1]{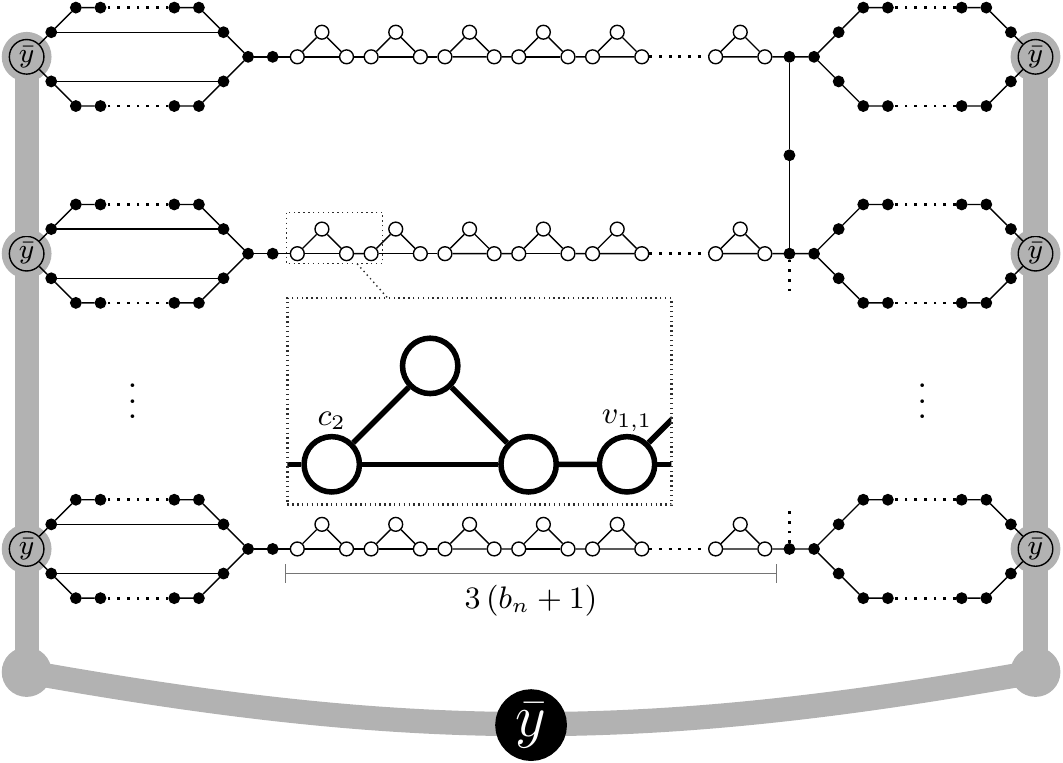}
			\caption{The graph $H$, containing a slightly different base-gadget as illustrated in Figure~\ref{fig:base_gadget_G}. All vertices connected by the thick gray line represent the same vertex $\bar y$. Circles represent $b$-paths containing $3(b_i + 1)$ vertices.}
			\label{fig:graph_h}
		\end{figure}

		The base-gadgets are part of the graphs $G$ and $H$ and are later used to characterize the MCS of those graphs. The graph $G$ represents the values of the elements in $X$ and $Y$, see Figure~\ref{fig:graph_g}. It contains $n$ \emph{$xy$-paths}, one between each pair of anchor vertices $c_i$ and $c_{i+n}$ for all $i\in\left[n\right]$. The $i$-th $xy$-path delineates the values of the elements $x_i$ and $y_i$. To this end, it consists of $s(x_i)$ connected $K_w^v$'s (called \emph{$x$-path}) which are connected to a path of length $2$ (called \emph{separating path}). In addition to the $x$-path, the separating path is also connected to $s(y_i)$ connected $K_w^v$'s (called \emph{$y$-path}).  Analogously, $H$ represents the values in the vector $\vec b$, see Figure~\ref{fig:graph_h}. There is a \emph{$b$-path} between each pair of anchor vertices $c_i$ and $c_{i+n}$  for all $i\in\left[n\right]$. Those $b$-paths represent the values of $\vec b$. For this purpose the $i$-th $b$-path consists of $b_i+1$ gadgets $K_w^v$'s.

		We now show that both graphs $G$ and $H$ are series-parallel and have degree bounded by $3$ for all but one vertex. We also show that both graphs can be computed in polynomial time with respect to the values of an NMwTS instance, which is necessary for the graphs to be used in a polynomial-time reduction.

\begin{figure}[t!]
			\centering
			\subfigure[]{
				\includegraphics[scale=1.1]{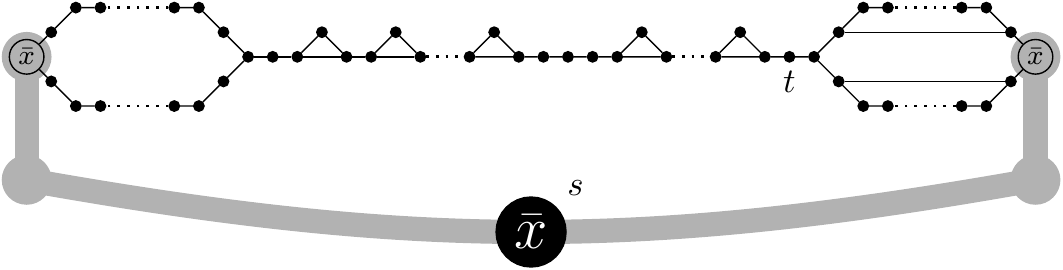}
				\label{sfig:construction_graph_G_a}
			}
			\subfigure[]{
				\includegraphics[scale=1.1]{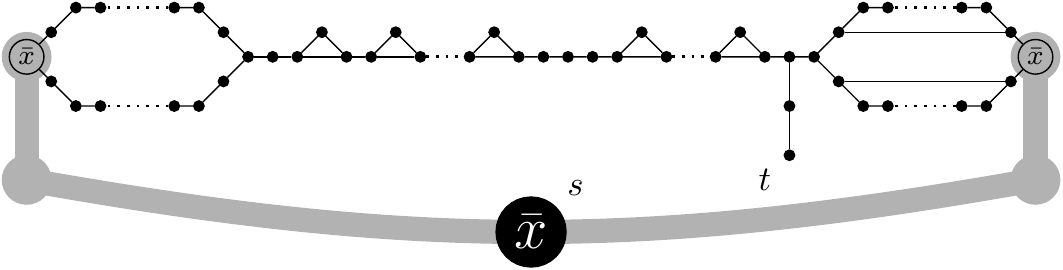}
				\label{sfig:construction_graph_G_b}
			}
			\subfigure[]{
				\includegraphics[scale=1.1]{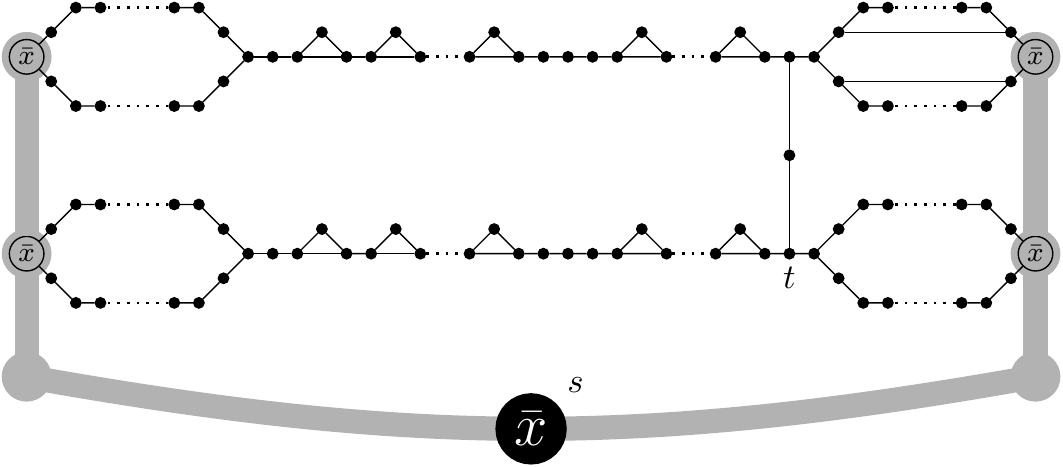}
				\label{sfig:construction_graph_G_c}
			}
			\caption{Construction of graph $G$ using $S$- and $P$-operations only. All vertices connected by the thick gray line represent the same vertex $\bar x$. \subref{sfig:construction_graph_G_a} Two cycles and an $xy$-path connecting two anchor vertices, which is a series-parallel graph. \subref{sfig:construction_graph_G_b} The same graph as before with an additional path attached to a prime anchor vertex which is still series-parallel. \subref{sfig:construction_graph_G_c} The graph obtained by a $P$-operation merging the $s$- and $t$-nodes of the graphs \subref{sfig:construction_graph_G_a} and \subref{sfig:construction_graph_G_b}.}
			\label{fig:sp_construction_of_G}
		\end{figure}

		\begin{lem}
			\label{lem:biconnected}
			The graphs $G$ and $H$ are biconnected.
		\end{lem}
		\begin{pf}
			Assume that $G$ is $1$-connected, but not $2$-connected. Thus, there is a vertex $v\in\Vertices{G}$ such that $v$ is a separator of $G$. Now there are three cases concerning the graph $G^{\prime}:=G\setminus \left\{ v \right\}$:
			\begin{enumerate}
				\item If $v$ is $\bar x$, then $G^{\prime}$ is still connected due to the anchor paths which connect all cycles containing chords which are again connected to the chordless cycles via the $xy$-paths.
				\item If $v\neq \bar x$ is contained in a cycle of the base-gadget (in a $C^{v}_{w}$ or $D^{v}_{w}$ gadget), then $G^{\prime}$ is still connected as cycles are biconnected and all cycles contain $\bar x$ as a common vertex.
				\item If $v$ is a vertex included in an anchor path or $xy$-path, then $G^{\prime}$ is still connected since each path is connected to two gadgets (of the same type, if $v$ is present in an anchor path, and of different type otherwise). Even if the path is split into two disjoint paths both of them remain connected to the base-gadget.
			\end{enumerate}
			In none of these cases $v$ is a separator and thus $G$ is biconnected as the cycles in the base-gadget are biconnected. The same arguments can be applied to $H$ if the $xy$-path is replaced with a $b$-path and the correct base-gadget is considered.\qed
		\end{pf}
		We now sketch a proof which shows that the graphs are series-parallel. The idea is to show that $G$ and $H$ can be constructed starting with a finite set of $K^{s,t}_2$ only using the $S$- and $P$-operations described in Sec.~\ref{sec:preliminaries}. The important steps of the construction are visualized in Figure~\ref{fig:sp_construction_of_G}.
		\begin{lem}
			The graphs $G$ and $H$ are series-parallel.
		\end{lem}
		\begin{pf}[Sketch]
			Paths and cycles are series-parallel. Hence, $K_w^v$'s are series-parallel and thus the $xy$-, $b$-paths and base-gadgets are series-parallel, too. They can be merged with $P$-operations such that $\bar x$ and an anchor vertex are the new $s$- and $t$-nodes. This process can be repeated until all $xy$- or $b$-paths are merged into their respective graphs.\qed
		\end{pf}
		Since NMwTS is \textbf{NP}-complete in the strong sense \cite{garey79}, we are allowed to construct MCS instances for reduction with a size polynomial with respect to the values of the NMwTS instance.
		\begin{lem}
			The graphs $G$ and $H$ can be constructed in polynomial time with respect to the values of the NMwTS instance.
		\end{lem}
		\begin{pf}
			We show that the number of vertices and edges in both graphs $G$ and $H$ is polynomial in the values of the NMwTS instance:
			Due to the construction, the size of both base-gadgets is equal. Hence, $\vert\Vertices{B_G}\vert=\vert\Vertices{B_H}\vert=4n\Ss+7n$. $G$ and $H$ also contain $xy$- and $b$-paths. Therefore, $\vert\Vertices{G}\vert=4n\Ss+7n+3\Ss+3n$ and $\vert\Vertices{H}\vert=4n\Ss+7n+3\Sb+3n$. Thus both sizes are polynomial in the values of an NMwTS instance. Since the degree is bounded by $3$ for all but one vertex in both graphs, the number of edges in both graphs is polynomial in those values, too. Consequently, these graphs can be constructed in polynomial-time.\qed
		\end{pf}

\begin{figure}[t!]
			\centering
			\subfigure[]{
				\includegraphics[scale=1.1777]{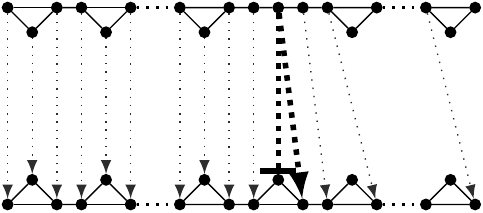}
				\label{sfig:path_matching_a}
			}\hfill
			\subfigure[]{
				\includegraphics[scale=1.1777]{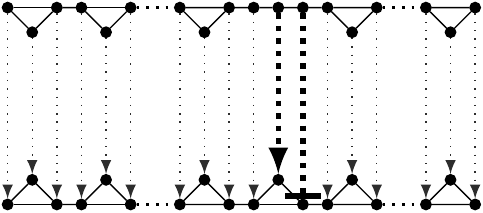}
				\label{sfig:path_matching_b}
			}
			\caption{Possible mappings of $xy$-path (top) to a $b$-path (bottom) indicated by dashed arrows. In \subref{sfig:path_matching_a} the thick line with the bar as tip is not a feasible mapping, the only possibility is the thick arrow. As a result only two vertices of each of the following $K_3$'s can be mapped. In \subref{sfig:path_matching_b} all the vertices of the $K_3$'s of the $x$- and $y$-paths are mapped. Only two of the vertices in the separating path can be contained in the mapping as the thick line with the bar as tip is not a feasible mapping.}
			\label{fig:matching_of_xy-_and_b-path}
		\end{figure}

		All MCS of $G$ and $H$ have common characteristics regarding their size and the vertices contained in them. First we show, that not all vertices in the $xy$- and $b$-paths can be contained in an MCS, see Figure~\ref{fig:matching_of_xy-_and_b-path} for an illustration.
		\begin{lem}
			\label{lem:mcs_of_paths}
			Let $P$ be an $xy$-path and $P^{\prime}$ be a $b$-path each with an additional edge incident to the vertices with degree one, then an MCS of $P$ and $P^{\prime}$ has size $\min\left(\vert\Vertices{P}\vert,\vert\Vertices{P^{\prime}}\vert\right)-1$.
		\end{lem}

		\begin{pf}
			Due to the construction there are $k,l\in\mathbb{N}$ such that $3k=\vert\Vertices{P}\vert$ and $3l=\vert\Vertices{P^{\prime}}\vert$. If $k\leq l$, then the $xy$-path contains at least one $K_3$ less than the $b$-path. Since the separating path cannot be mapped to a $K_3$ there is at least one vertex which cannot be contained in an MCS. If $k>l$, then the $xy$-path contains at least two more $K_3$'s than the $b$-path, hence each vertex except one in the $b$-path can be contained in the MCS.\qed
		\end{pf}

		Since $G$ and $H$ only contain the base-gadget and $n$ $xy$- or $b$-paths, we also have to consider the base-gadget in an MCS of $G$ and $H$. We can show, that all vertices in the base-gadgets except the ones in the anchor paths are contained in the MCS.
		\begin{lem}
			\label{lem:mapping_of_base_gadgets}
			Let $B_{G}$ and $B_{K}$ be the base-gadgets of $G$ and $H$ and $\bar x$ and $\bar y$ be the vertices whose degree is not bounded by $3$ in $B_G$ and $B_H$, respectively. Then an MCS of $B_{G}$ and $B_{H}$ has size $\vert\Vertices{B_{G}}\vert-n+1$ and $\bar x$ is mapped to $\bar y$.
		\end{lem}
		\begin{pf}
			Note that $\vert\Vertices{B_{G}}\vert=\vert\Vertices{B_{H}}\vert$. We first show that the vertices with unbounded degree ($\bar x$ in $B_G$ and $\bar y$ in $B_H$) are mapped to each other and then determine the size of an MCS. Assume that $\bar x$ is not mapped to $\bar y$ but to $v\in\Vertices{B_H}$ such that $v\neq\bar y$ in a common subgraph $F$. Due to the construction, the degree of $v$ is at most $3$ and $\bar x$ has degree $4n$. Therefore, not all of the cycles containing $\bar x$ ($C^{u}_{w}$ and $D^{u}_{w}$ type gadgets) can be contained in $F$. If $\bar x$ and $\bar y$ are mapped to each other, all cycles can be contained in a common subgraph, thus $F$ cannot be an MCS.

			Now let $F$ be a common subgraph of $B_G$ and $B_H$ where $\bar x$ is mapped to $\bar y$. In this case we only have to consider how the cycles of the base-gadgets are contained in $F$, i.e., how are the vertices in the cycles mapped in $F$. If a chordless cycle is mapped to a cycle containing chords, the following vertices cannot be contained in $F$: the anchor, prime anchor and the other vertex the prime anchor vertex is adjacent to in both cycles. If cycles are split up, i.e., one cycle is mapped to two other cycles, then there is at least one vertex per cycle which cannot be contained in $F$. Both cases result in at least $n$ vertices which are not contained in $F$. If however all cycles are mapped according to their gadget type, then only the vertices in the anchor paths cannot be in $F$. There are $n-1$ such vertices. Therefore, if all cycles are mapped that way, $F$ is an MCS of size $\vert\Vertices{B_G}\vert-n+1$.\qed
		\end{pf}

	
	\subsection{Correctness of the Polynomial-Time Reduction} 
	\label{sub:the_polynomial_time_reduction}
	We have shown that $G$ and $H$ can be computed in polynomial time and that an MCS of $G$ and $H$ has some characteristics regarding its size and the vertices it contains.	Concerning the polynomial-time reduction, we show that an instance of NMwTS has a numerical matching if and only if an MCS of the corresponding graphs $G$ and $H$ has a specific size.
		\begin{lem}
			\label{lem:mcs_of_nmwts_instance}
			An instance $\left(X,Y,s,\vec b\right)$ of NMwTS has a numerical matching if and only if $\vert\Vertices{G}\vert=\vert\Vertices{H}\vert$ and an MCS of $G$ and $H$ has size $\left\vert\Vertices{G}\right\vert-2n+1$.
		\end{lem}

		\begin{pf}
			Let $\left(X,Y,s,\vec b\right)$ be an instance of NMwTS and $G,H$ graphs constructed as described above. Assume that there is a numerical matching. Hence, $\Sigma_s=\Sigma_{\vec b}$ and thus $\vert\Vertices{G}\vert=\vert\Vertices{H}\vert$. An MCS of all $xy$- and $b$-paths has size $\left\vert\Vertices{G}\right\vert-n$ according to Lemma~\ref{lem:mcs_of_paths} and an MCS of the base-gadgets has size  $\left\vert\Vertices{G}\right\vert-n+1$, see Lemma~\ref{lem:mapping_of_base_gadgets}. Even though they have been considered separately, the results can be combined, since all relevant vertices (the ones adjacent to the base-gadget and the ones in $xy$-paths and $b$-paths) are contained in each MCS.

			Now assume $\vert\Vertices{G}\vert=\vert\Vertices{H}\vert$ and there is an MCS with size $\left\vert\Vertices{G}\right\vert-2n+1$. Since we only consider connected MCSs, the vertex with unbounded degree must be contained in this MCS. For each $xy$-path and $b$-path there has to be one vertex which cannot be contained in an MCS (Lemma~\ref{lem:mcs_of_paths}). The same is true for the base-gadgets, since the vertices of the anchor paths cannot be contained (Lemma~\ref{lem:mapping_of_base_gadgets}). The vertices of the separating paths are not contained in an MCS. Thus the values of the elements of $X$ and $Y$ are correctly bipartitioned. Due to the size of the graphs for each $b_i$ there is an $x_j$ and $y_{j^\prime}$ such that $b_i=s(x_j)+s(y_{j^\prime})$. \qed
		\end{pf}
		Since $G$ and $H$ both have a maximum degree bounded by $3$ for all but one vertex, we obtain the following result.
		\begin{thm}
			\label{the:np_hardness_of_mcs}
			\mcsb{3}{1} in biconnected series-parallel graphs is \textbf{NP}-hard.
		\end{thm}


\section{Block-and-Bridge Preserving MCS in Series-Parallel Graphs} 
\label{sec:the_block-and-bidge_preserving_maximum_common_subgraph_problem_in_series-parallel_graphs}
	In this section we consider the task of finding a maximum common subgraph that preserves the structure of blocks and bridges of the input graphs (BBP-MCS). This problem has been introduced in \cite{Schietgat2007} and is also discussed in \cite{Akutsu2012}.
   A common subgraph is said to be \emph{block-and-bridge preserving} (BBP) if
	\begin{axioms}{BBP}
		\item any two vertices in different blocks in the common subgraph must not be contained in the same block of an input graph, and \label{ax:bbp:blocks}
		\item each bridge in the common subgraph is a bridge in both input graphs.\label{ax:bbp:bridge}
	\end{axioms}
	In order to solve this problem, we make use of the following observation:
	Every two vertices in an input graph that are not in the same block cannot be in the same block in any common subgraph. 
	Hence, together with conditions (\ref{ax:bbp:blocks}) and (\ref{ax:bbp:bridge}) we may infer that vertices in one block of $G$ can only be mapped to vertices contained in exactly one block of $H$ such that the resulting common subgraph is biconnected. This means that blocks and bridges can be considered separately to some extent and allows us to solve BBP-MCS by systematically decomposing the input graphs and solving the associated subproblems. We use the BC-tree data structure to decompose the connected input graphs into tree-like parts consisting of bridges and maximal biconnected subgraph, i.e., the blocks. Every block is then again decomposed by means of SP-trees. Solving MCS for the bridges that appear in the BC-tree is similar to the problem $1$-MCS while the mapping between the blocks is closely related to $2$-MCS. For this, we use the ideas of a recently proposed polynomial-time algorithm based on the SP-tree data structure \cite{Kriege2014a}.
	Clearly, whenever two vertices of blocks are mapped that are both cutvertices of the input graphs, the components beyond these cutvertices must be considered for BBP-MCS.

	\subsection{Decomposing Series-Parallel Graphs}
\begin{figure}
		\centering
		\subfigure[Input graph $G$]{
			\includegraphics[scale=.95]{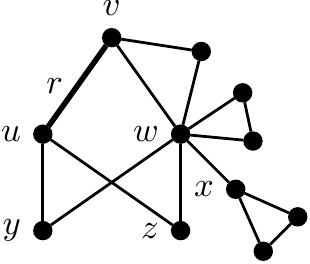}
			\label{fig:bc:graph}
		}\hfill
		\subfigure[BC-tree $T^{\BC}_G$]{	
			\includegraphics[scale=.95]{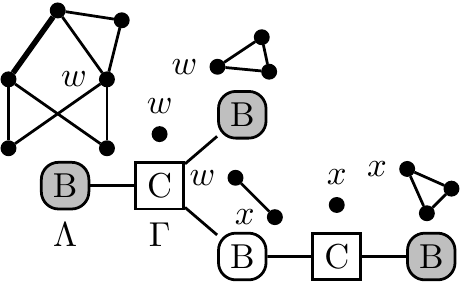}
			\label{fig:bc:graph_bc-tree}
		}\hfill
		\subfigure[SP-tree $T^{\SP}_\Lambda$]{	
			\includegraphics[scale=.95]{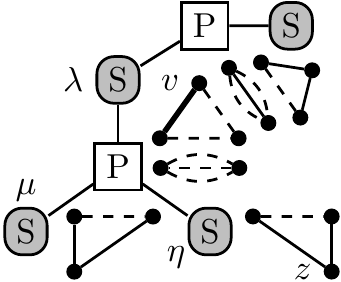}
			\label{fig:bc:graph_sp-tree}
		}
		\caption{\subref{fig:bc:graph} A series-parallel graph and \subref{fig:bc:graph_bc-tree} its BC-tree with the associated skeleton graphs. Block $B$-nodes have a gray background color, while bridge $B$-nodes are not filled. \subref{fig:bc:graph_sp-tree} The SP-tree of the skeleton graph associated with the block $B$-node $\Lambda$ together with its skeleton graphs.}
		\label{fig:bc}
	\end{figure}

	We consider simple connected series-parallel graphs and decompose them by means of BC- and SP-tree data structures. We refer to the vertices of SP- and BC-trees as \emph{nodes} to distinguish them from the vertices of the input graphs.  We use a notation and definition as introduced in \cite{Chimani2011} simplified and slightly modified for our application to series-parallel graphs.
	Let $G$ be a simple connected graph, recall that a block is a maximal biconnected subgraph of $G$ and a bridge is an edge that is not contained in a block. Hence, any two blocks and bridges of $G$ may have at most one vertex in common, which must be a cutvertex. 
	\begin{defn}[BC-tree]
		Given a connected graph $G$ with at least two vertices, let $C$ denote the set of cutvertices, $Bl$ the set of blocks, $Br$ the set of bridges and $B = Br \cup Bl$. The \emph{BC-tree} $T^{\BC}=\BC(G)$ of $G$ is the tree with nodes $B \cup C$ and edges between nodes $b \in B$ and $c\in C$ iff $c\in\Vertices{b}$.
	\end{defn}
	We refer to the nodes of $T^{\BC}$ representing cutvertices as $C$-nodes and distinguish between bridge and block $B$-nodes; given a BC-tree $T^\BC$, we refer to the three sets of nodes by $V_{\C}(T^\BC)$, $V_{\Br}(T^\BC)$ and $V_{\Bl}(T^\BC)$, respectively. Each node $\Lambda$ in a BC-tree has a skeleton graph $S_{\Lambda}$ consisting of the vertices and edges of $G$ represented by that node, i.e., a single vertex for $C$-nodes, a $K_2$ for bridge $B$-nodes and a biconnected subgraph for block $B$-nodes, cf. Figure~\ref{fig:bc}. 

	SP-trees allow to decompose biconnected series-parallel graphs and can consequently be applied to the skeleton graphs of block $B$-nodes.

	\begin{defn}[SP-tree]\label{def:sp_tree}
		Let $G$ be a biconnected series-parallel graph with at least three vertices. 
		The \emph{SP-tree} $T^{\SP}=\SP(G)$ is the smallest tree that satisfies the following properties:
		\begin{axioms}{SP}
			\item Each node  $\lambda$ of $T^{\SP}$ is associated with a \emph{skeleton} graph $S_{\lambda}=(V_{\lambda},E_{\lambda})$. Each edge $e=(u,v)\in E_{\lambda}$ is either a real or a virtual edge. If $e$ is a virtual edge, then $S=\left\{ u,v \right\}$ is a separator of $G$.\label{def:sp_tree:skeleton}
			\item $T^{\SP}$ has two different node types with the following skeleton
				structures:
				\begin{description}[topsep=.3em,parsep=0pt,itemsep=.3em,leftmargin=1.6em]
					\item[S:] The skeleton $S_\lambda$ is a simple chordless cycle, i.e., $\lambda$ represents a series composition.
					\item[P:] The skeleton $S_\lambda$ consists of two vertices and multiple parallel edges between them, i.e., $\lambda$ represents a parallel composition.
				\end{description}
			\item For two adjacent nodes $\lambda$ and $\eta$ in $T^{\SP}$, the skeleton graph $S_{\lambda}$ contains a virtual edge $e_{\eta}$ that represents $S_{\eta}$ and vice versa. The node $\eta$ is called \emph{pertinent} to the edge $e_{\eta}$.
			\item The graph resulting by merging all skeleton graphs in a way that each virtual edge is replaced by the skeleton of its pertinent node is exactly $G$.\label{def:sp_tree:merge}
		\end{axioms}
	\end{defn}
The sets of $S$-nodes and $P$-nodes in $T^{\SP}$ are denoted by $V_S(T^{\SP})$ and $V_P(T^{\SP})$, respectively. The SP-tree $T^{\SP}$ is bipartite with respect to these two sets. Since a vertex $v$ of a graph may occur in multiple skeleton graphs of the SP-tree $T^\SP$, we denote by $\lambda(v)$ the representative of $v$ in the skeleton $S_\lambda$. For the sake of simplicity we do not distinguish vertices of the original graph and their representatives in skeleton graphs, when this is clear from the context.
Let $r \in \Edges{G}$, the SP-tree \emph{rooted at $r$} is obtained by rooting $T^{\SP}$ at the node $\lambda \in \Vertices{T^{\SP}}$ such that $r$ is a real edge in $S_{\lambda}$. 
A rooted SP-tree induces a parent-child relation, where a node $\lambda \in \Vertices{T^{\SP}}$ is the parent of an adjacent node $\eta \in \Vertices{T^{\SP}}$ if the path from the root node to $\lambda$ is shorter than the path from the root node to $\eta$. If a node $\lambda \in \Vertices{T^{\SP}}$ is the parent of a node $\eta \in \Vertices{T^{\SP}}$ and $\lambda$ is the  node pertinent to an edge $e_{\lambda} \in \Edges{S_{\eta}}$, then $e_{\lambda}$ is called a \emph{reference edge} of $\lambda$ and denoted by $\REF(\lambda)$.
For the readers convenience, Greek upper- and lower-case letters denote the $B$-, $C$- and $S$-, $P$-nodes of the graph decompositions, respectively. Latin letters denote the vertices of the input graphs.
Figure~\ref{fig:bc:graph_sp-tree} shows an example of an SP-tree and the skeleton graphs associated with the individual nodes. It is well known that BC- as well as SP-trees can be computed in linear time~\cite{Gutwenger2001}.

	\begin{figure}
		\centering
		\null\hfill
		\subfigure[$G^r_{vw}$]{
			\includegraphics{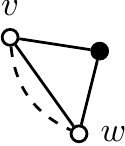}
			\label{fig:split:1}
		}\hfill
		\subfigure[$G^r_{uw}$]{	
			\includegraphics{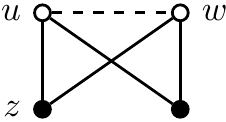}
			\label{fig:split:3}
		}\hfill
		\subfigure[$G^r_w$]{	
			\includegraphics{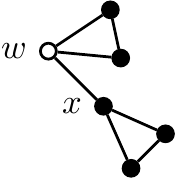}
			\label{fig:split:2}
		}\hfill
		\subfigure[$G^r_{vz}$]{	
			\includegraphics{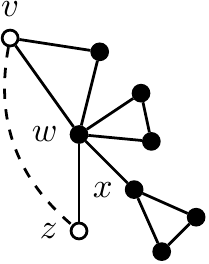}
			\label{fig:split:4}
		}\hfill\null
		\caption{Some split graphs obtained from the graph depicted in Figure~\ref{fig:bc:graph}: In \subref{fig:split:1} and \subref{fig:split:3} the graph is split at different $2$-separators, in \subref{fig:split:2} at a cutvertex and in \subref{fig:split:4} at a potential separator. Base vertices are not filled.}
		\label{fig:split}
	\end{figure}

Our BBP-MCS algorithm computes the solution based on subproblems for well-defined subgraphs of the input graphs closely related to these two data structures.
Given a set of vertices $S$, let $C^S_i$ denote the connected components of $G \setminus S$
and assume w.l.o.g. that at least one endpoint of a distinguished edge $r$ is contained in $C^S_1$. Let $\mathcal{C}^r_S = \bigcup_{i \geq 2} V(C^S_i)$. 
We consider the following \emph{split graphs}:
Let $S=\{v\}$ with $v$ a cutvertex, then $G^r_v$ denotes the graph $G[\mathcal{C}^r_S \cup S]$ with \emph{base vertex} $v$.
Let $S=\{u,v\}$ be a minimal separator, then $G^r_{uv}$ denotes the graph 
$G[\mathcal{C}^r_S \setminus (\mathcal{C}^r_{\{u\}} \cup \mathcal{C}^r_{\{v\}}) \cup S]$; we say $u$ and $v$ are the base vertices of $G^r_{uv}$.
Figure~\ref{fig:split} shows an example of several split graphs obtained from the graph depicted in Figure~\ref{fig:bc:graph}.
Let $e=(u,v)$ be an edge that is part of a block. If $e \neq r$, then $G^r_{uv}$ is defined as the edge $e$ itself, otherwise it is the graph $(V, E \setminus e)$. 

Note that there is a one-to-one correspondence between split graphs obtained from cutvertices and specific \emph{rooted subtrees} of the BC-tree of $G$. Let $G^r_v$ be a split graph, consider the $C$-node $\Gamma$ representing the base vertex $v$ and the unique $B$-node $\Lambda$ with $r \in E(S_\Lambda)$. The tree rooted at $\Gamma$, where the child subtree containing $\Lambda$ is deleted, is associated with the split graph. For example, the split graph depicted in Figure~\ref{fig:split:2} corresponds to the subtree of the BC-tree shown in Figure~\ref{fig:bc:graph_bc-tree} that is obtained by rooting the tree at $\Gamma$ and deleting $\Lambda$.

For handling block $B$-nodes we build on the algorithm described in \cite{Kriege2014a}, which computes a maximum common biconnected subgraph of two series-parallel graphs. Again, the split graphs we consider are closely related to the data structure we use for the decomposition, i.e., SP-trees.
The set $S=\{u,v\}$ is a (minimal) $2$-separator of a biconnected graph if and only if both vertices are contained in the skeleton graph of an $S$-node and are not connected by a real edge.
Since $2$-separators are not sufficient to obtain correct solutions in biconnected series-parallel graphs, the concept of \emph{potential separators} has been introduced. These are not only the $2$-separators of the graph $G$, but the $2$-separators of any induced biconnected subgraph of $G$. In a series-parallel graphs all pairs of non-adjacent vertices that are contained in a chordless cycle are potential separators. For example, in Figure~\ref{fig:bc:graph} the set $S=\{v,z\}$ is a potential separator, but not a separator of the graph, since $u$ remains connected to $w$ in $G \setminus S$ through the vertex $y$. Consequently, this vertex cannot be contained in the subgraph that is separated by $S$. Removing these components systematically as detailed in \cite{Kriege2014a} until a potential separator becomes a separator of the graph then allows us to apply our notion of split graphs also in this case, cf. Figure~\ref{fig:split:4}. These split graphs can be obtained by means of SP-trees, where one vertex of a potential separator is contained in a skeleton graph of an $S$-node $\lambda$ and the other in an $S$-node $\eta$. Removing all branches starting from $P$-nodes on the unique path with endpoints $\lambda$ and $\eta$ corresponds to the maximal biconnected subgraph that is separated by $S$. Consider the SP-tree shown in Figure~\ref{fig:bc:graph_sp-tree}, the vertex $v$ is contained in $S_\lambda$ and $z$ in $S_\eta$ and the component associated with the node $\mu$ consequently is removed to obtain a subgraph, which is then split to obtain $G^r_{vz}$.

\SetKwFunction{BBPMCS}{BbpMcs}
\SetKwFunction{BBPMCSS}{Series}
\SetKwFunction{MATCHEDGE}{Edge}
\SetKwFunction{MATCHVERTEX}{Cut}
\SetKwFunction{MWBM}{MwbMatching}
\SetKwFunction{ROOT}{Root}
\SetKwFunction{NEXT}{Next}

\subsection{The Algorithm} 
\label{sub:the_algorithm}
We now present a polynomial-time algorithm, which solves BBP-MCS in series-parallel graphs based on the decomposition of input graphs by means of BC- and SP-trees. We apply the idea presented in \cite{Matula1978} for MCS in trees to the BC-trees constructing a solution based on smaller subproblems between rooted subtrees, which are combined with MwbM. In a similar manner the block $B$-nodes are handled by means of SP-trees, where certain additional technicalities must be considered~\cite{Kriege2014a}.

\begin{algorithm}[tb]
  \caption[]{$\BBPMCS{G,H}$}\label{alg:mcs}
  \Input{Two series-parallel graphs $G$ and $H$.}
  \Output{Size of a BBP-MCS of the graphs $G$ and $H$.}
  \Data{BC-trees $T^{\BC}_G$ and $T^{\BC}_H$ of $G$ and $H$, respectively,
        SP-trees $T^{\SP}_{\Lambda}$ for every block $B$-node $\Lambda$ in $T^{\BC}_G$ and 
        $T^{\BC}_H$.}

  $mcs \gets 0$ \;
  \ForAll(\note*[f]{Pairs of block $B$-nodes})
  {$(\Lambda,\Lambda^{\prime}) \in V_{\Bl}(T^{\BC}_G) \times V_{\Bl}(T^{\BC}_H)$} { \label{alg:mcs:blocks:start}
    $\ROOT(T^{\BC}_G,\Lambda)$; $\ROOT(T^{\BC}_H,\Lambda^{\prime})$ \;
    \ForAll(\note*[f]{Pairs of $S$-nodes})
    {$(\lambda,\lambda^{\prime})\in V_S(T^{\SP}_{\Lambda})\times V_S(T^{\SP}_{\Lambda^{\prime}})$}{
      $r \gets$ arbitrary $(u,v)\in E(S_{\lambda})\cap E(G)$;
      $\ROOT(T^{\SP}_\Lambda,r)$ \;
      \ForAll{edges $r^{\prime}=(u^{\prime},v^{\prime})\in E(S_{\lambda^{\prime}})\cap E(H)$} {
        $\ROOT(T^{\SP}_{\Lambda^{\prime}},r^{\prime})$ \;
        $p \gets \BBPMCSS(u,v,u^{\prime},v^{\prime})$ \;
        $p \gets \BBPMCSS(u,v,v^{\prime},u^{\prime})$ \note*[r]{Alternative edge mapping}
        $mcs \gets \max\{mcs, p, q\}$ \; \label{alg:mcs:blocks:end}
      }
    }
  }
  \ForAll(\note*[f]{Pairs of bridge $B$-nodes})
  {$(\Lambda,\Lambda^{\prime}) \in V_{\Br}(T^{\BC}_G) \times V_{\Br}(T^{\BC}_H)$} { \label{alg:mcs:bridges:start}
    $\ROOT(T^{\BC}_G,\Lambda)$; $\ROOT(T^{\BC}_H,\Lambda^{\prime})$ \;
    $r=(u,v) \gets E(S^{\BC}_{\Lambda})$;
    $r^{\prime}=(u^{\prime},v^{\prime}) \gets E(S^{\BC}_{\Lambda^{\prime}})$  \;
    $p \gets \MATCHVERTEX(u,u^{\prime})+\MATCHVERTEX(v,v^{\prime})$ \;
    $q \gets \MATCHVERTEX(u,v^{\prime})+\MATCHVERTEX(v,u^{\prime})$ \note*[r]{Alternative edge mapping}
    $mcs \gets \max\{mcs, p, q\}$ \; \label{alg:mcs:bridges:end}
  }
  \Return $\max\{1, mcs+2\}$ \label{alg:mcs:return}\;
\end{algorithm}

The main procedure \BBPMCS is presented as Algorithm~\ref{alg:mcs}. Given two series-parallel graphs as input, we assume that the associated BC- and SP-trees are already computed, which can be done in linear time. The algorithm computes the size of a BBP-MCS between the two input graphs by recursively dividing the problem into smaller subproblems defined between split graphs. The main procedure starts with pairs of $B$-nodes in the BC-trees of the input graphs. Since blocks and bridges must be preserved, only pairs of block $B$-nodes (Line~\ref{alg:mcs:blocks:start}--\ref{alg:mcs:blocks:end}) and pairs of bridge $B$-nodes (Line~\ref{alg:mcs:bridges:start}--\ref{alg:mcs:bridges:end}) are considered. 

In the former case, the algorithm loops over all possible pairs of $S$-nodes in the SP-trees of the two blocks. Two distinguished edges serve as starting point of the mapping by fixing the mapping of their endpoints. The resulting subproblem then is handled by the procedure \BBPMCSS, see Algorithm~\ref{alg:bbp-mcs-s}. The main procedure does not take into account that the vertices $u$ and $u'$ (or $v$ and $v'$) may be cutvertices or that the edges $r$ and $r'$ could be virtual. These cases are not considered since the possible better result would be obtained anyway for a different starting configuration.

In the latter case, the edges associated with the bridge $B$-nodes serve as starting point of the mapping. Again, the two possible mappings of endpoints are considered and the solution is determined based on the return values of \MATCHVERTEX, see Algorithm~\ref{alg:bbp-mcs-c}. Assume we have the bridges $r=(u,v)$ of $G$ and $r'=(u',v')$ of $H$ and map the vertex $u$ to $u'$ and $v$ to $v'$. The result is obtained by combining the BBP-MCS between the split graphs $G^r_u$ and $G^{r'}_{u'}$ with the BBP-MCS between the opposing split graphs $G^r_v$ and $G^{r'}_{v'}$.

Finally, the procedure returns the best solution found and accounts for the two base vertices, which have been mapped, by adding two. It is possible that no solution has been found, e.g., when one graph consists of bridges only and the other merely of blocks. Note that a single vertex can always be mapped without violating \eqref{ax:bbp:blocks} and \eqref{ax:bbp:bridge}. Hence, in this case the value $1$ is returned (Line~\ref{alg:mcs:return}), where we assume the input graphs to be non-empty. 

For the computation of the BBP-MCS between two split graphs, we require that the mapping of base vertices is fixed and compute the maximum possible solution under this constraint. Whenever a procedure is called for this purpose the BC-trees and SP-trees considered have been rooted by the procedure \ROOT at distinct $B$-nodes and at distinct edges, respectively. Therefore, we can make use of the parent-child relationship between adjacent nodes. In particular, we may infer which parts of the graphs have already been considered, since these are associated with the branches containing the root. We will give the details of the procedures called by \BBPMCS in the following.

\begin{algorithm}[tb]
  \caption[]{$\BBPMCSS(u,v,u^{\prime},v')$}\label{alg:bbp-mcs-s}
  \Input{Base vertices $u,v$ of $G$, $v \in V(S_\lambda)$ and $u', v'$ of $H$, $v' \in V(S_{\lambda'})$.}
  \Output{Size of a BBP-MCS between $G^r_{uv}$ and $H^{r'}_{u'v'}$ 
          under the restriction that $u$ is mapped to $u'$ and $v$ to $v'$.}
  $e = (v,w) \gets \NEXT(v, S_\lambda)$ \note*[r]{Next edge in $S_\lambda$} \label{alg:mcs-ser:simple:start}
  $e' = (v',w') \gets \NEXT(v', S_{\lambda'})$ \note*[r]{Next edge in $S_{\lambda'}$}

  \lIf(\note*[f]{Parent $S$-node}){$e = \REF(\lambda)$}   { \Return $\BBPMCSS(u, \parentS(v), u', v')$ } \label{alg:mcs-ser:simple:merge1}
  \lIf(\note*[f]{Parent $S$-node}){$e' = \REF(\lambda')$} { \Return $\BBPMCSS(u, v, u', \parentS(v'))$ } \label{alg:mcs-ser:simple:merge2}
  \If(\note*[f]{Completed skeleton}){$w=u$ \KwAnd $w'=u'$} { \Return $\MATCHEDGE(v,w,v',w')$ } \label{alg:mcs-ser:simple:complete}
  \lIf(\note*[f]{Incompletable mapping}){$w=u$ \KwXor $w'=u'$} { \Return $-\infty$ } \label{alg:mcs-ser:simple:incomp}

  $mcs \gets \MATCHEDGE(v,w,v',w') + \MATCHVERTEX(w,w') + \BBPMCSS(u, w, u', w') + 1$ \label{alg:add_bbp-mcs-c_2}\label{alg:mcs-ser:simple:end}\;

  \If(\note*[f]{Consider potential separators}){$e \notin E(G)$ \KwOr $e' \notin E(H)$} { \label{alg:mcs-ser:merge:start}
    \lIf{$e \in E(G)$} { \label{alg:mcs-ser:merge:start2}
      $M \gets \{\lambda\}$
    } \lElse {
      $M \gets \childS(e)$\label{alg:mcs-ser:merge:child1}
    }
    \lIf {$e' \in E(H)$} {
      $M' \gets \{\lambda'\}$
    } \lElse {
      $M' \gets \childS(e')$\label{alg:mcs-ser:merge:child2}
    } 
    \ForAll{$(\eta,\eta') \in M \times M'$} { \label{alg:mcs-ser:merge:forall}
      $p \gets \BBPMCSS(u, \eta(v), u', \eta'(v'))$ \note*[r]{Continue in child $S$-node} 
      $mcs \gets \max\{mcs, p\}$ \label{alg:mcs-ser:merge:end}\;
    }
  } 
  \Return $mcs$ \;
\end{algorithm}

The procedure \BBPMCSS essentially computes the $2$-MCS between two blocks based on their SP-trees as in \cite[$\textsc{Mcs-S}$]{Kriege2014a}, but is modified to take cutvertices into account. For a virtual edge $e$ in the skeleton $S_\lambda$ of an $S$-node $\lambda$ we denote the children of the $P$-node pertinent to $e$ by $\childS(e)$. For a vertex $v$ that is one endpoint of the reference edge in a skeleton $S_\mu$ we refer to the representative of $v$ in the next $S$-node on the path to the root by $\parentS(v)$.
The function $\NEXT(u,S_\lambda)$ returns the vertex adjacent to $u$ which yet has not been considered in the skeleton graph of $\lambda$.
To obtain a BBP-MCS the common subgraph of two blocks must be biconnected. Hence, we simultaneously traverse the cyclic skeleton graphs of two $S$-nodes, one of $T^\SP_G$ and the other of $T^\SP_H$, starting from one endpoint of an edge in each. Eventually, we have to return to the other endpoints of these edges in both graphs at the same time. In each step the mapping is extended by the next vertex $w$ lying on the cycle starting from the second base vertex $v$. In the case that the edge $(v,w)$ is virtual, the edge represents parts of the graph beyond this edge, which are considered by the procedure \MATCHEDGE. If the next vertex $w$ is a cutvertex in the input graph, the parts of the graphs that are not contained in the block are taken into account by the procedure \MATCHVERTEX. Lines~\ref{alg:mcs-ser:merge:start} to \ref{alg:mcs-ser:merge:end} handle potential separators, where the two base vertices are contained in different $S$-nodes. 

We illustrate the procedure and how the approach aligns with our definition of split graphs in the following.
Consider the example shown in Figure~\ref{fig:bc:graph} and assume that the main procedure \BBPMCS has started with a pair of block $B$-nodes, where the first is $\Lambda \in V(T^\BC_G)$. Further, assume that the $S$-node $\lambda$ has been selected from the SP-tree associated with the block, which has been rooted at the edge $r$. Then, the next vertex in $S_\lambda$ is $w$, which is connected to the base vertex $v$ by the virtual edge $(v,w)$. The situation corresponds to the case that is handled in Line~\ref{alg:add_bbp-mcs-c_2} of the procedure \BBPMCSS. The problem here is divided into three subproblems: First, if possible, the part of the input graph represented by the virtual edge must be mapped, i.e., the split graph $G^r_{vw}$, see Figure~\ref{fig:split:1}. This is done by the procedure \MATCHEDGE, explained in detail later. Second, the vertex $w$ is a cutvertex in $G$ and the split graph $G^r_w$, see Figure~\ref{fig:split:2}, must be taken into account whenever $w$ is mapped to a cutvertex in $H$. This is handled by the procedure \MATCHVERTEX. Finally, the remainder of the skeleton $S_\lambda$ must be considered, which is done by the recursive call of \BBPMCSS, where the vertex $w$ replaces $v$ as base vertex. Consequently, this procedure tries to map the split graph $G^r_{uw}$, see Figure~\ref{fig:split:3}. Since the vertex $w$ is mapped, one is added to the sum over the return values of these procedures.

\begin{algorithm}[tb]
  \caption[]{\MATCHEDGE{$u,v,u',v'$}}\label{alg:mcs-e}
  \Input{Vertices with $e=(u,v) \in E(S_\lambda)$ and $e'=(u',v') \in E(S_{\lambda'})$.}
  \Output{Size of a BBP-MCS of $G^r_{uv}$ and $H^{r'}_{u'v'}$ under the 
          restriction that $u$ is mapped to $u'$ and $v$ to $v'$.}

  \lIf(\note*[f]{Subgraph not induced}){$e \in E(G)$ \KwXor $e' \in E(H)$} { 
    \Return $-\infty$ \label{alg:mcs-par:matching:non-induced}
  } 
  \lIf(\note*[f]{Valid mapping}){$e$ is real in $S_\mu$ \KwOr $e'$ is real in $S_{\mu'}$}{
    \Return $0$ \label{alg:mcs-par:matching:valid}
  }
  $M \gets \childS(e)$; \label{alg:mcs-par:matching:start} \label{alg:mcs-par:matching:child1} 
  $M' \gets \childS(e')$ \; \label{alg:mcs-par:matching:child2}
  \ForAll(\note*[f]{Pairs of $S$-node children}){$m=(\eta,\eta') \in M \times M'$} {
    $w(m) \gets \BBPMCSS(\eta(u), \eta(v), \eta'(u'), \eta'(v'))$ \; \label{alg:mcs-par:matching:weights}
  }
  $p \gets \MWBM(M,M',w)$ \note*[r]{Compute maximum weight matching} \label{alg:mcs-par:matching:end}
  \lIf{$p \neq 0$ \KwOr $e \in E(G)$, $e' \in E(H)$ } {
    \Return $p$ \label{alg:mcs-par:matching:return2}
  } \lElse(\note*[f]{Not biconnected}){
    \Return $-\infty$ \label{alg:mcs-par:matching:return1}
  }
\end{algorithm}

The two procedures \MATCHEDGE and \MATCHVERTEX both construct MwbM instances to determine the matching between unmapped connected components obtained from a $2$-separator and a cutvertex, respectively. Algorithm~\ref{alg:mcs-e} gives the details of \MATCHEDGE, which is called with the endpoints of two edges as arguments, one in the skeleton of the $S$-node $\lambda$ and the other in the skeleton of $\lambda'$.
Note that it is possible that there is a virtual edge $(u,v)$ in a skeleton graph, although $(u,v)$ is not an edge of the input graph. If only one of the edges is contained in the input graph, the mapping is not allowed since we would not obtain a common induced subgraph.
If both edges are virtual, then the endpoints are separators, see (\ref{def:sp_tree:skeleton}), and the BBP-MCS of the split graph has to be added to the result. This is done by computing the maximum weight matching in the complete bipartite graph $C$, where the two vertex sets are the children of the $P$-node pertinent to $e$ and $e'$, respectively.
The weight  $w\colon E(C)\to \mathbb{N}\cup \left\{ -\infty \right\}$ of an edge $e=(\lambda,\lambda')$, where $\lambda \in V_S(T^\SP_G)$ and $\lambda' \in V_S(T^\SP_H)$, is the size of a BBP-MCS between the two components associated with its endpoints. 
This value is determined by the procedure \BBPMCSS starting from the reference edges of the $S$-nodes $\lambda$ and $\lambda'$. Note that if the result of the matching is $0$ and there is no edge between the base vertices, then the BBP-MCS between the considered split graphs does not contain a path connecting the base vertices. Since in this case the common subgraph would not be biconnected, contradicting (\ref{ax:bbp:bridge}), the value $-\infty$ is returned (Line~\ref{alg:mcs-par:matching:return1}).

\begin{algorithm}[tb]
  \caption[]{$\MATCHVERTEX(u,u')$}\label{alg:bbp-mcs-c}
  \Input{Two vertices $u \in V(G)$, $u' \in V(H)$.}
  \Output{Size of a BBP-MCS between $G^r_u$ and $H^{r'}_{u'}$ 
          under the restriction that $u$ is mapped to $u'$; $0$ if $u$ and $u'$ not both 
          are cutvertices.}
  \lIf{\text{$u$ and $u'$ \KwNot both are cutvertices}} {
    \Return $0$
  }
  $M \gets \childB(u)$; $M' \gets \childB(u')$ \note*[r]{Get $B$-node children}
  \ForAll{$m=(\Lambda,\Lambda')\in M \times M'$} {
    \uIf{$\Lambda$ and $\Lambda'$ both are bridge $B$-nodes} {
      $(u,v) \gets E(S_\Lambda)$; $(u',v') \gets E(S_{\Lambda'})$ \note*[r]{Get associated bridges}
      $w(m) \gets \MATCHVERTEX(v,v') +1$ \;
    } \uElseIf{$\Lambda$ and $\Lambda'$ both are block $B$-nodes} {
      \note{Get the $S$-nodes containing an edge incident to base vertices}
      $N \gets \{\lambda \in V_S(T^{\SP}_{\Lambda}) \mid \exists r = (u,v) \in E(S_{\lambda}) \cap E(G) \}$ \;
      $N' \gets \{\lambda' \in V_S(T^{\SP}_{\Lambda'}) \mid \exists r' = (u',v') \in E(S_{\lambda'}) \cap E(H) \}$ \;
      \ForAll(\note*[f]{Pairs of relevant $S$-nodes})
      {$(\lambda,\lambda') \in N \times N'$}{
        $r \gets$ arbitrary $(u,v)\in E(S_{\lambda}) \cap E(G)$;
        $\ROOT(T^{\SP}_{\Lambda},r)$ \;
        \ForAll {edges $r'=(u',v')\in E(S_{\lambda'}) \cap E(H)$} {
          $\ROOT(T^{\SP}_{\Lambda'},r^{\prime})$ \;
          $w(m) \gets \max\{w(m), \BBPMCSS(u,v,u',v')\}$ \;
        }
      }
    } \lElse(\note*[f]{Non BBP matching}) { $w(m) \gets -\infty$ \label{alg:bbp-mcs-c:non-bbp}}
  }
  \Return $\MWBM(M,M',w)$ \;
\end{algorithm}

The procedure \MATCHVERTEX computes the size of a BBP-MCS of two split graphs obtained from cutvertices. Therefore, zero is returned if the given vertices $u$ and $u'$ are not both cutvertices. Otherwise, we consider their child $B$-nodes $\childB(u)$ and $\childB(u')$ in the rooted BC-trees. 
To this end, we again create a weighted complete bipartite graph $C$ with vertex partition $\childB(u) \cup \childB(u^{\prime})$. The weight $w\colon E(C)\to \mathbb{N}\cup \left\{ -\infty \right\}$ of an edge is the size of a BBP-MCS of the two split graphs associated with its endpoints. Note that the node sets contain block $B$-nodes as well as bridge $B$-node. Edges connecting different types of $B$-nodes obtain weight $-\infty$ (Line~\ref{alg:bbp-mcs-c:non-bbp}) as mapping them contradicts restriction (\ref{ax:bbp:blocks}). This assures that these edges are not contained in any maximum weight matching.
The weight of an edge between two bridge $B$-nodes is determined by a recursive call of \MATCHVERTEX, where the current base vertices are replaced by the other endpoints of the bridges.
In case of two block $B$-nodes the BBP-MCS is determined in a similar manner as in the main procedure, with the difference that only $S$-nodes are considered that contain a representative of the base vertex with an appropriate incident edge $r \in E(G)$, which is mapped to an edge $r' \in E(H)$. These two edges serve as roots of the SP-trees and for each pair the procedure $\BBPMCSS$ is called.
The maximum value returned by any of these calls yields the edge weight. Note that the mapping of the base vertices is fixed and---in contrast to the main procedure---we do not need to consider all the possible mappings.

\subsection{Analysis} 
\label{sub:analysis}
We analyze Algorithm~\ref{alg:mcs} and show that BBP-MCS can be solved in polynomial time. We give improved bounds on the running time for the case that both input graphs are outerplanar.
\begin{thm}
	\label{the:runningtime_bbp}
	The problem BBP-MCS in series-parallel graphs can be solved in time $\mathcal{O}(n^6)$, where $n$ is the number of vertices in the larger input graph.
\end{thm}

\begin{pf}
	The correctness of the algorithm is based on the argumentation above and the results of~\cite{Kriege2014a}. To prove the running time, we transform the algorithm in a dynamic programming approach. In \cite[Theorem 1]{Kriege2014a} it is shown, that $2$-MCS can be solved in time $\mathcal{O}(n^6)$ when storing the $2$-MCS between all pairs of split graphs in a table of size $\mathcal{O}(n^4)$. Note that we also have to consider split graphs that are obtained from cutvertices, for which another table of size $\mathcal{O}(n^2)$ is sufficient.

	We consider the running time required by the procedure \BBPMCSS. Assume that for each split graph that is smaller than the split graph considered by the current call, the BBP-MCS has been computed. Then all calls to procedures are answered in constant time by a lookup in the table and the running time for calling \BBPMCSS once is $\mathcal{O}(n^2)$. Since in each call a cell of the table is filled the total running time caused by all calls of this procedure is $\mathcal{O}(n^6)$.

	The procedures \MATCHVERTEX and \MATCHEDGE both require to solve MwbM problems. In case of \MATCHVERTEX, each matching problem corresponds to a pair of rooted subtrees of the BC-trees. There can be at most $\mathcal{O}(n^2)$ such pairs and each matching problem can be solved in $\mathcal{O}(n^3)$ by the Hungarian method. This results in a total running time of $\mathcal{O}(n^5)$. For the procedure \MATCHEDGE, each matching problem corresponds to a pair of subtrees of the SP-trees. Each subtree is obtained for some P-node, where one adjacent S-node is removed. With the same argument as above the total running time of the procedure is $\mathcal{O}(n^5)$ as in~\cite{Kriege2014a}.
   Finally, we consider the main procedure \BBPMCS. Since we may assume that all calls to procedures are answered in constant time by lookups in the tables, the running time of the procedure is $\mathcal{O}(n^2)$. Consequently, the total running time of the algorithm is dominated by the procedure \BBPMCSS and is $\mathcal{O}(n^6)$. \qed
\end{pf}
If we restrict to outerplanar graphs, each $P$-node in the SP-trees has degree $2$ which concludes in the following theorem.
\begin{thm}
	BBP-MCS in outerplanar graphs can be solved in time $\mathcal{O}\left(n^5\right)$.
\end{thm}
\begin{pf}
	The proof is similar to the proof of Theorem \ref{the:runningtime_bbp}. Since all $P$-nodes in SP-trees of outerplanar graphs have degree $2$, the total running time of \BBPMCSS reduces to $\mathcal{O}(n^4)$, since the sets $M$ and $M'$ considered in Line~\ref{alg:mcs-ser:merge:forall} contain only one element.
	Moreover, there is no need to use MwbM in the procedure \MATCHEDGE as the bipartite graphs are $K_2$'s. Consequently the running time of a single call of \MATCHEDGE becomes constant.
	For the procedure \MATCHVERTEX the restriction to outerplanar graphs does not allow improved bounds on the running time since the number of rooted subtrees of the BC-tree does not change. Therefore, the total running time is $\mathcal{O}(n^5)$.\qed
\end{pf}

It is known that BBP-MCES in outerplanar graphs can be solved in polynomial time \cite{Schietgat2007,Schietgat2008,Schietgat2013,Akutsu2013}, where MCES refers to a variation of the problem that asks for edge-induced common subgraphs with maximum number of edges. Note that---in contrast to the variant we consider---a subgraph where in one input graph two vertices are adjacent while the vertices in the other are not, is an MCES, but not an MCS.
In \cite{Akutsu2013} the rough bound of $\mathcal{O}(n^{10})$ for a BBP-MCES algorithm was given. The approach by Schietgat et al. has been shown to be efficient in practice~\cite{Schietgat2013}. Different worst-case bounds are provided in several publications starting with $\mathcal{O}(n^7)$~\cite{Schietgat2007}, then $\mathcal{O}(n^5)$~\cite{Schietgat2008} and finally $\mathcal{O}(n^2 \sqrt{n})$~\cite{Schietgat2013}.


\section{Concluding Remarks} 
\label{sec:concluding_remarks}
	We showed that MCS in series-parallel graphs with degree bounded by $3$ for all but one vertex is \textbf{NP}-hard by reduction from NMwTS. Moreover, we extended a $2$-MCS algorithm \cite{Kriege2014a} to solve BBP-MCS with running time $\mathcal{O}(n^6)$. In outerplanar graphs, it can solve BBP-MCS in $\mathcal{O}(n^5)$. BBP-MCES in outerplanar graphs was taken as basis to obtain polynomial time solutions for MCES in outerplanar graphs of bounded degree~\cite{Akutsu2012}. It is still unknown whether MCS or MCES in series-parallel graphs is solvable in polynomial time if all vertices have bounded degree. To the authors' best knowledge, there is only one problem which is known to be solvable in polynomial time in outerplanar graphs, but is \textbf{NP}-complete in series-parallel graphs: the \emph{edge-disjoint paths problem} \cite{Nishizeki2001}. Since series-parallel graphs are equivalent to the partial $2$-trees, there is a parameterized class of graphs, i.e., the partial $k$-trees, for which it is known that MCS is \textbf{NP}-hard for $k\geq 11$ even when the degree is bounded~\cite{Akutsu2013}. For all other $k > 1$, the complexity has yet to be decided.

\section*{Acknowledgement}
We would like to thank the anonymous reviewers for their valuable comments and suggestions to improve and clarify this manuscript.

\bibliographystyle{elsarticle-num}
\bibliography{lit}

\end{document}